\documentclass{PoS}

\usepackage{amsmath}

\title{Witten index from lattice simulation}

\ShortTitle{Witten index from lattice simulation}

\author{\speaker{Issaku Kanamori}%
         \thanks{
The current affiliation: 
Institut f\"ur Theoretische Physik, Universit\"at Regensburg,
D-93040 Regensburg, Germany
}
\\
INFN sezione di Torino, and Dipartimento di Fisica Teorica, 
Universit\`a di Torino,\\
Via P. Giuria 1, 10125 Torino, Italy\\

        E-mail: \email{issaku.kanamori@physik.uni-regensburg.de}}

\abstract{
I propose a method for measuring the Witten index using a lattice
simulation.  The index is useful to discuss spontaneous breaking of
supersymmetry.  As a test of the method, I also report some 
numerical results for the supersymmetric quantum mechanics, for which the
index is known.}

\FullConference{The XXVIII International Symposium on Lattice Filed Theory\\
                 June 14-19,2010\\
                 Villasimius, Sardinia Italy}

\newcommand{\Tr}{\mathop{\rm tr}\nolimits}

\newcommand{\psibar}{{\overline{\psi}}}

\newcommand{\Slash}[1]{%
{%
\setlength{\unitlength}{1em}%
\thicklines
\begin{picture}(0,1)
 \put(-.2,-.2){\line(#1,1){#1}}
\end{picture}%
}%
}

\begin{document}

\section{Introduction}
\label{sec:intro}

The supersymmetry (SUSY) is believed as a symmetry of the unification
theory such as superstring theory and supersymmetric gauge 
theory is a candidate of a theory beyond the Standard model.
However, it is broken in our current universe anyway.
Since it cannot be broken by higher loop effects in perturbation,
it is important to study the breaking nonperturbatively.

Witten index \cite{Witten:1982df}
is a useful index related to the spontaneous SUSY breaking,
which is defined nonperturbatively.
Using the fermion number operator $F$, it is given by the following trace,
\begin{equation}
 w=\Tr(-1)^F e^{-\beta H}= (N_B-N_F)\bigr|_{E=0}\ , 
\end{equation}
where $H$ is the Hamiltonian of the system and $E$ is its eigenvalue.  
As long as the spectrum
is discrete, the index does not depend on a parameter $\beta$.\footnote{
If the spectrum is continuum, one has to take a limit $\beta\to\infty$.
}
It is simply a difference of numbers of bosonic supersymmetric 
vacua and fermionic vacua.
If the index is not zero, there exists at least one supersymmetric
vacuum so SUSY is not broken.  But if the index is zero, 
SUSY may or may not be broken, since
it can be a
result of cancellation between bosonic and fermion vacua, or 
a result of no supersymmetric vacua at all. 
The purpose of this talk is to propose a method to measure the Witten
index using lattice simulation based on Ref.~\cite{Kanamori:2010gw}.
For a different approach from lattice simulation, see Ref.~\cite{Kawai:2010yj}.

In terms of the path integral, the index becomes a partition function
with periodic boundary condition \cite{Cecotti:1981fu,Fujikawa:1982nt}
\begin{equation}
  w=Z_{\rm P}=\int \mathcal{D\phi}\, \mathcal{D}\psibar\,\mathcal{D}\psi
   \exp(-S_{\rm P}),
\end{equation}
where $\phi$ is boson, $\psi$ and $\psibar$ are fermion,
and subscript $\rm P$ stands for periodic boundary conditions
for all the fields in the temporal direction.
It seems difficult to measure this quantity using lattice simulation,
since what we usually measure is an expectation value normalized by the
partition function but we need the normalization factor here.
The normalization of the path integral measure is relevant as well.

In the following section, we will discuss how to obtain the correct
normalization of the partition function and thus the Witten index.
And then in section~\ref{sec:numerical} 
we confirm that it in fact works in supersymmetric quantum mechanics 
of which the Witten index is well known using a lattice simulation.  
We also test a method which would improve the efficiency 
of the measuring the Witten index.

\section{Idea}
\label{sec:idea}

We have to determine two normalizations: one for the path integral
measure and the other is for the partition function (from the lattice data).

\begin{figure}
{
\hfil
{\setlength{\unitlength}{0.11em}
 \begin{picture}(150,60)(0,0)
  \put(0,-23){
  \includegraphics[width=150\unitlength]{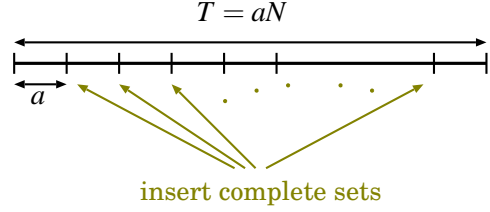} 
  }
  \put(11,30){\makebox(0,0){$a$}}
  \put(75,57){\makebox(0,0){$T=aN$}}
 \end{picture}
}

\caption{Obtaining the correct measure:  Derivation of the path integral
is exactly the lattice regularization.}
\label{fig:operator2pathingetral}
}
\end{figure}

The path integral measure with a correct normalization is 
easy to obtain.  We only have to follow a standard derivation
of the path integral from the operator formalism, where we insert
\emph{normalized} complete sets at each of discretized time slices (
Fig.~\ref{fig:operator2pathingetral}
).
Regarding the discretization is the lattice discretization, 
we obtain the following measures for bosons and fermions:
\begin{align}
&\text{Bosons:}&
  \int \mathcal{D}\phi 
 &= \int_{-\infty}^{\infty}\prod_i\frac{1}{\sqrt{2\pi}}d\phi_i^{\rm(lat)},
    && \label{eq:measure-b}\\
&\text{Fermions:}&
 \int\mathcal{D}\psibar\, \mathcal{D}\psi
     &= \int \prod_i d\psibar_i^{\rm (lat)}\, d\psi_i^{\rm (lat)}.
 \label{eq:measure-f}
\end{align}

The correct normalization of the partition function is non-trivial.
Let us start with a 1-dimensional 
bosonic system with $N$ lattice sites 
and consider the following quantity:
\begin{equation}
 \left\langle e^{+S} {
  \underbrace{\textstyle
 \exp\left[-\frac{1}{2}\sum\limits_i\mu^2 (\phi_i^{\rm lat})^2\right]}_{%
  \text{regularization functional}}} \right\rangle
 \equiv
  \frac{C}{\int \mathcal{D}\phi\, e^{-S}},
 \label{eq:observable}
\end{equation}
where $\mu$ is an arbitrary (positive) real number which should be tuned
later and
\begin{equation}
 C= \int { \mathcal{D}\phi}\, 
 \exp\left[-\frac{1}{2}\sum\limits_i\mu^2 (\phi_i^{\rm lat})^2\right]
  = \mu^{-N}.
 \label{eq:C}
\end{equation}
Here, we have used eq.~(\ref{eq:measure-b}).
Combining eq.~(\ref{eq:observable}) and (\ref{eq:C}), we obtain
\begin{equation}
 Z=\int \mathcal{D}\phi\, e^{-S}
  =\frac{C}{\left\langle \exp\left[+S-\frac{1}{2}\sum\limits_i\mu^2 (\phi_i^{\rm lat})^2\right] \right\rangle}.
 \label{eq:partition-function}
\end{equation}
Since we can calculate the value of $C$ analytically,
and the denominator in the r.h.s is an observable in the lattice
simulation, we can measure
the partition function $Z$.
Notice that though we have used a gaussian functional as a regularization
functional in eq.~(\ref{eq:observable}), one can use any functional
as long as it gives a calculable and convergent value like in eq.~(\ref{eq:C}).

In the r.h.s.\ of eq.~(\ref{eq:partition-function}), 
the action $S$ appears with a ``wrong sign'' which cancels the original
distribution.   That is, the partition function is calculated using an
extreme reweighting.
To obtain a better efficiency, we have to tune the value of $\mu$.

Next let us introduce fermions.
After integrating out the fermions, we obtain the effective action
as usual:
\begin{equation}
 S' = S_{\rm B} - \ln |\det (D)|,
\end{equation}
where $S_{\rm B}$ is the bosonic part of the action and $D$ is the
the fermion bilinear operator (i.e., the Dirac operator plus the Yukawa
interactions)\footnote{If the fermion is Majorana, the determinant 
should be replaced with a Pfaffian}.
The phase factor of $\det(D)$ should be reweighted afterwards which gives
for arbitrary expectation values
\begin{equation}
 \langle A \rangle
   = \frac{\int \mathcal{D}\phi\, A \sigma[D] e^{-S'}}
          {\int \mathcal{D}\phi\, \sigma[D] e^{-S'}}
   = { \frac{\langle A\sigma[D]\rangle_0}{\langle \sigma[D] \rangle_0} } ,
 \label{eq:phase-reweight}
\end{equation}
where $\sigma[D]$ is the phase factor and the subscript $0$ stands for a phase
quenched average.
This time we have to cancel e a factor $\sigma[D]e^{-S'}$ to obtain the
partition function.  
Therefore, measuring
$
\langle \sigma[D]^{-1}e^{+S'} \exp(-\frac{1}{2}\sum_i \mu^2 (\phi_i^{\rm
lat})^2)\rangle
$,
we obtain the Witten index as
\begin{equation}
 w=Z_{\rm P}= C\frac{\langle \sigma[D_{\rm P}] \rangle_{0,{\rm P}}}
  {\left\langle \exp\left[+S'_{\rm P}-\frac{1}{2}\sum_i \mu^2 \phi_i^2\right]\right\rangle_{0, {\rm P}}
},
\label{eq:w}
\end{equation}
where $C$ is given in eq.~(\ref{eq:C}).

The r.h.s of eq.~(\ref{eq:phase-reweight}) implies the phase quenched
average of the phase factor $\sigma[D]$ is almost the partition
function.  This observation is correct, and eq.~(\ref{eq:w}) provides
the correct normalization to the partition function.

\section{Numerical Test: Supersymmetric Quantum Mechanics}
\label{sec:numerical}

We test our method using supersymmetric quantum mechanics (of
$\mathcal{N}=2$ Wess-Zumino type) \cite{Witten:1981nf}, 
of which the Witten index is known.

If the lattice action keeps a part of supersymmetry as an exact symmetry
on the lattice, we expect that the Witten index is well defined.
More precisely, if the action is given as $Q\Lambda$ with 
an exact supertransformation which satisfies $Q^2=0$, we can repeat a
similar argument to the continuum case.  As a result, the index
is well defined even at finite lattice spacing in such lattice models.
In particular, the index from a finite lattice spacing should be an integer.

A $Q$-exact lattice action for the supersymmetric quantum mechenics
is given as \cite{Catterall:2000rv}
\begin{equation}
 S
  =\sum_{k=0}^{N-1}\Bigl[
    \frac{1}{2}(\phi_{k+1}-\phi_k)^2 + \frac{1}{2}W'(\phi_k)^2
     +(\phi_{k+1}-\phi_k)W'(\phi_k) -\frac{1}{2}F_k^2 
    + \psibar_k(\psi_{k+1} -\psi_k) + W''(\phi_k)\psibar_k\psi_k
    \Bigr],
\end{equation}
where $\phi_k$ is a real boson, $\psi_k$ and $\psibar_k$ are fermions,
and $F_k$ is a real bosonic auxiliary field.
The potential $W$ is a function of $\phi$ 
and the prime ($'$) indicates a derivative.
If the asymptotic behavior is $W(+\infty)W(-\infty)>0$ the supersymmetry
is not broken and $W(+\infty)W(-\infty)<0$ it is broken.  We use the
following two cases:
\begin{itemize}
 \item $n=4$: $W=\lambda_4 \phi^4 + \lambda_2 \phi^2$ 
       \qquad SUSY, \quad $w=1$
 \item $n=3$: $W=\lambda_3 \phi^3 + \lambda_2 \phi^2$
       \qquad \Slash{3}SUSY, \quad $w=0$
\end{itemize}
where $\lambda_i$ are parameters of the potential.
We use the Hybrid Monte Carlo algorithm.  See \cite{Kanamori:2007yx}
for the implementation for this system.

The results are plotted in Figs.~\ref{fig:n4} and \ref{fig:n3}.
With a suitable choice of $\mu^2$, the known indexes are reproduced.
There is almost no dependence on the lattice spacing, as expected from
the exact $Q$-symmetry of the action.

\FIGURE{

 \includegraphics[width=0.49\linewidth]{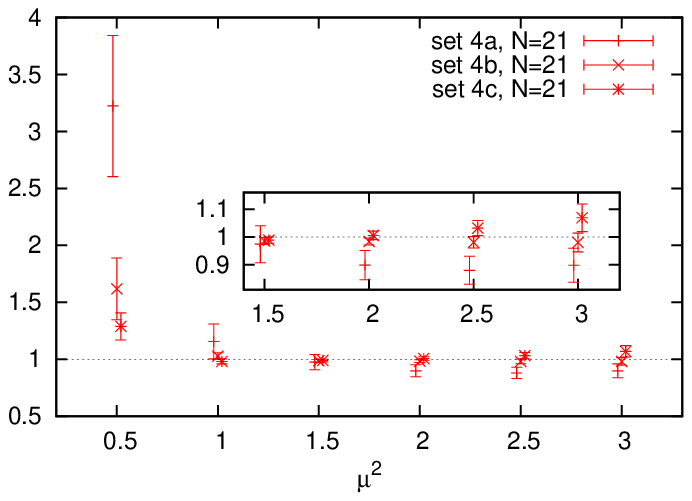}
 \hfil
 \includegraphics[width=0.49\linewidth]{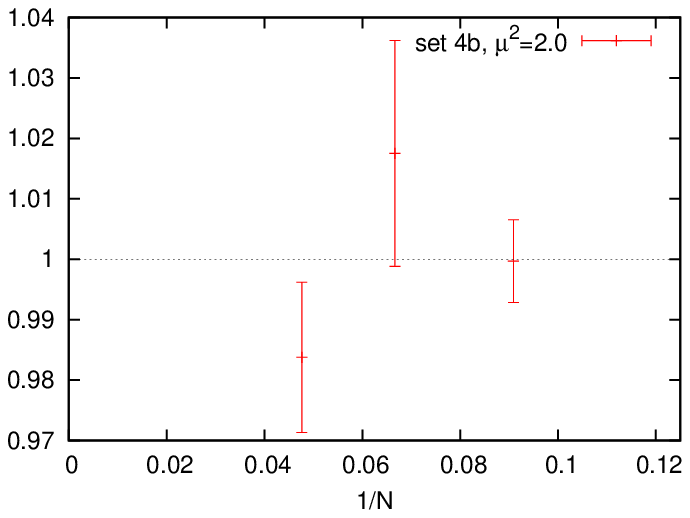}

\parbox{0.8\linewidth}{
set 4a ($L \lambda_2 =1, L^2 \lambda_4=1$) : $\mu^2=2.5,\ w=0.88(5)$\\
set 4b ($L \lambda_2 =4, L^2 \lambda_4=1$) : $\mu^2=2.0,\ w=0.984(12)$\\
set 4c ($L \lambda_2 =4, L^2 \lambda_4=4$) : $\mu^2=1.5,\ w=0.989(11)$
}

 \caption{$n=4$ case, where the index is known to be $1$.
 $L$ is the physical size of the system.
 (left panel) $\mu^2$ dependence.  (right panel) Lattice spacing $a=1/N$
 dependence.  (bottom) values of $\mu^2$ and the measured index $w$, which
 minimize the error.
}
 \label{fig:n4}
}

\FIGURE{

 \includegraphics[width=0.49\linewidth]{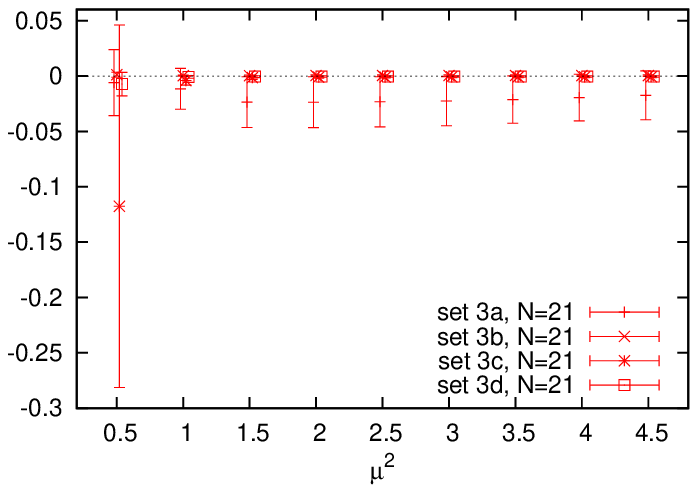}
 \hfil
 \includegraphics[width=0.49\linewidth]{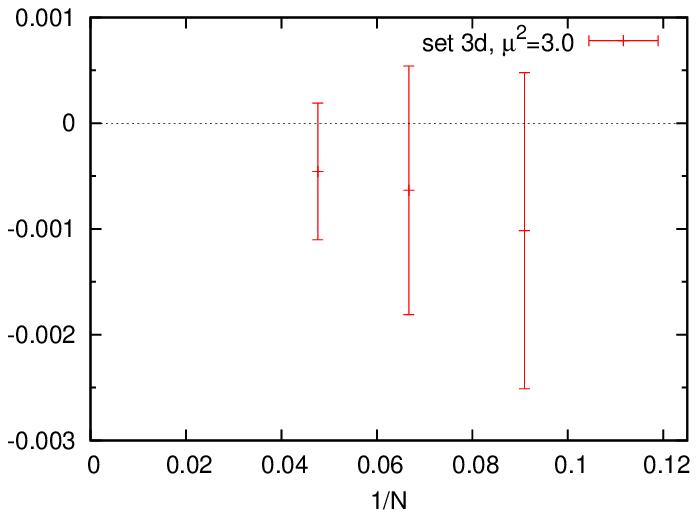}

\parbox{0.8\linewidth}{
set 3a ($L \lambda_2 =4, L^{3/2} \lambda_4=4\ $) : $\mu^2=1.5, -0.024(23)$\\
set 3b ($L \lambda_2 =4, L^{3/2} \lambda_4=16$)  : $\mu^2=2.0, 0.0004(7)$\\
set 3c ($L \lambda_2 =4, L^{3/2} \lambda_4=32$)  : $\mu^2=1.5, -0.0009(8)$\\
set 3d ($L \lambda_2 =2, L^{3/2} \lambda_4=16$)  : $\mu^2=1.5, -0.0005(6)$
}

\caption{$n=3$ case, where the index is known to be $0$.
 $L$ is the physical size of the system.
 (left panel) $\mu^2$ dependence.  (right panel) Lattice spacing $a=1/N$
 dependence.  (bottom) values of $\mu^2$ and the measured index $w$, which
 minimize the error.
}
 \label{fig:n3}
}

Next, we consider a possible way to improve the efficiency.
Because of the factor $e^{S'_{\rm P}}$ in eq.~(\ref{eq:w}), 
the efficiency is poor and we need large statistics.
This factor cancels the weight from the action so we do not have to use
importance sampling with respect to a weight factor $e^{-S}$.  
Therefore, we can also use configurations generated with 
\emph{less} importance sampling.
Decomposing the weight factor
as $e^{-S}=e^{-rS}e^{-(1-r)S}$, we rewrite a general expectation value
as
\begin{equation}
 \langle A \rangle
  = \frac{\int \mathcal{D}\phi\, { Ae^{-rS}} e^{-(1-r)S} }
    {\int \mathcal{D}\phi\, { e^{-rS}} e^{-(1-r)S} }
  = \frac{\langle { Ae^{-rS}}\rangle_r}{\langle { e^{-rS}}\rangle_r},
\end{equation}
where $\langle\ \cdot\ \rangle_r$ is an expectation value with a weight
factor $e^{(1-r)S}$.  Therefore, preparing configurations using
$e^{(1-r)S'_{\rm P}}$, we can obtain the Witten index as follow:
\begin{equation}
  w = C 
    \frac{\langle \sigma[D_{\rm P}] { e}^{ -rS'_{\rm P}}\rangle_{r,{\rm P}}}
         {\langle { \exp}{\left[ { (1-r)S'_{\rm P}}
          -\frac{1}{2}\sum_i  \mu^2(\phi_i^{\rm lat})^2 \right]}
          \rangle_{r,{\rm P}}}.
\end{equation}
Note that $r=0$ is the usual importance sampling.

We plot the result from the less importance sampling in Fig.~\ref{fig:li}.
On the left panel, we see that the correct index is reproduced with a
suitable choice of $\mu^2$.
On the right panel, we plot the behavior of the errors versus number of
the configurations used in the measurements.
Contrary to the naive expectation, the magnitudes of the error are the
same for large statistics in both $r=0$ case and $r>0$ case.  
For small statistics, however, $r>0$ cases converge to a line
$(\text{num. of confs.})^{-1/2}$ faster than $r=0$ case.
This implies that the less importance sampling method is robuster
for small statistics.

\FIGURE
{

\includegraphics[width=0.49\linewidth]{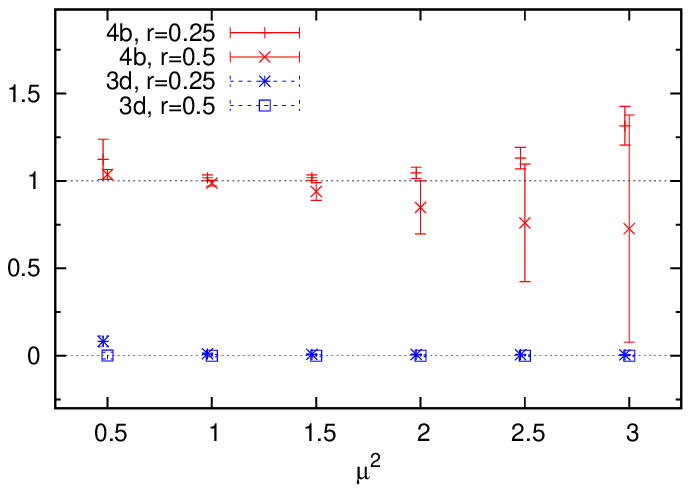}
\includegraphics[width=0.49\linewidth]{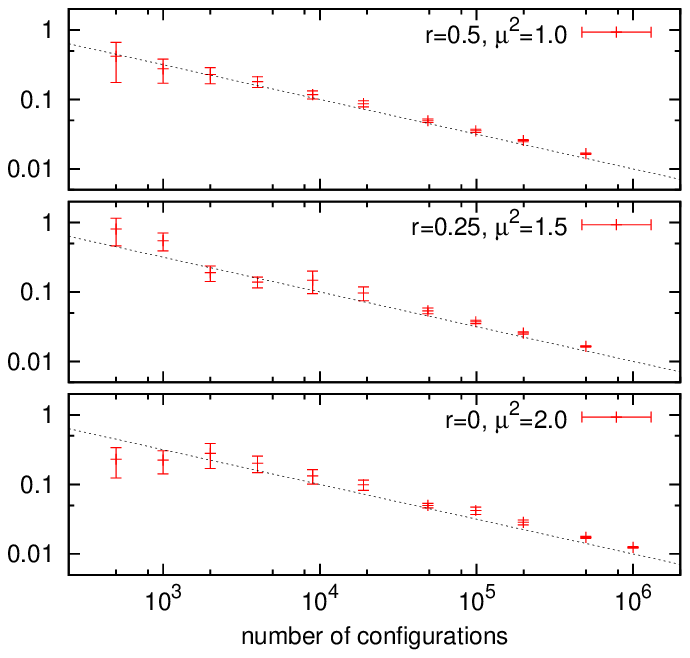}
\caption{Results from the less importance sampling. (left panel)
 The obtained Witten index. Set labels are the same in Figs.~\ref{fig:n4}
 and \ref{fig:n3}.
 (right panel) Behavior of the errors, for set 4b. The dotted line is
$(\text{num. of conf.})^{-1/2}$.}
\label{fig:li}
}

\section{Conclusion and Discussion}
\label{sec:conclusion}

We proposed a method for measuring the Witten index, which is a useful
index to detect a spontaneous supersymmetry breaking.
Since the index is given as a partition function under the periodic
boundary condition, it is important to use the correct normalization of
the path integral measure.
We also normalized overall factor of the partition function measuring a
special regularization functional.
As a test of the method, we measured the index of supersymmetric quantum
mechanics.  The results reproduced the known values of the index.
A disadvantage of the method is its poor efficiency.  A less importance
sampling method may improve it to some extend.

Finally, we mention possible applications of the method, which may or may
not be practical.
It is straightforward to use the method in higher dimensional systems.
Within one-dimensional systems, the most interesting one is supersymmetric
Yang-Mills quantum mechanics with 16 supercharges.  This model is one of
the candidates of M(atrix)-theory, and assumes the Witten index should
be 1 to obtain a suitable supergravity limit.

\subsection*{Acknowledgements}
I. K. was financially supported by Nishina Memorial
Foundation.
He also thanks Insituto Nazionale di Fisica Nucleare (INFN).

\end{document}